\documentstyle[graphicx,aps]{revtex}
\newcommand{\tc}{$T_{\rm c}$}
\newcommand{\srt}{SrTiO$_3$}
\newcommand{\Dq}[1]{$\Delta$q$_{\rm #1}$}
\newcommand{\ang}[1]{$\times$10$^{#1}$\AA$^{-1}$}
\newcommand{\vc}[1]{\mathbf{#1}}
\def\e{{\rm e}}

\begin{document}
%
%
\preprint{\sf Phys. Rev. B}
\title{
Influence of defects on the critical behaviour 
at the \boldmath{$105$}\,K structural phase transition of SrTiO$_3$:
I. The broad component} 
\author{H.~H\"unnefeld, T.~Niem\"oller, and J.~R.~Schneider} 
\address{Hamburger Synchrotronstrahlungslabor HASYLAB at Deutsches
Elektronen-\\
Synchrotron DESY, Notkestr. 85, D-22603 Hamburg, Germany}
\author{B.~A.~Kaufmann and F.~Schwabl}
\address{Lehrstuhl f\"ur Theoretische Physik V, Physik-Department der 
Technischen Universit\"at  M\"unchen, \\
James-Franck-Stra\ss e, D-85747 Garching, Germany}
\date{June 7, 2000}
\maketitle
\begin{abstract}
  The critical fluctuations in SrTiO$_3$ near its $105\,K$ structural
  phase transition were studied with triple axis diffractometry using
  high energy ($\ge 100\,keV$) synchrotron radiation in different
  SrTiO$_3$ crystals with different oxygen vacancy concentrations.
  Due to the presence of oxygen vacancies the critical behaviour is
  changed compared to defect-free systems.  In our experiments a
  smearing out of the squared order parameter and a crossover of the
  critical exponents $\nu$ and $\gamma$ above the phase transition
  temperature is observed, with the crossover temperature strongly
  depending on the concentration of the defects.  To understand the
  experimental findings, e.g.~the unusual values for the critical
  exponents found near the critical temperature, the
  Ginzburg-Landau-Wilson functional for structural phase transitions
  in disordered systems was analyzed using renormalization group
  theory and the replica trick.  Considering the effects of defects
  which locally increase the transition temperature leads to a
  qualitative understanding of the observed behaviour.  The crossover
  behaviour of the critical exponents can be modeled and a
  quantitative analysis of the observed experimental data is
  presented.
\end{abstract}
\newpage 
\section{Introduction}
The influence of defects on the critical behaviour close to a
structural phase transition has been studied both experimentally and
theoretically for a long time. These investigations show that the
critical behaviour can change dramatically compared to homogeneous
systems. According to the Harris' criterium the critical behaviour is
modified by quenched disorder if the critical exponent $\nu$ of the
correlation length obeys the inequality $\nu > 2/d$ with $d$ the
spatial dimensionality of the system \cite{Har74a}. As a result the
critical exponents near the phase transition temperature can change
their values \cite{Khm75,Gri76}. In addition a smearing out of the
specific heat can be observed \cite{Hal76,Str80}. Furthermore the
central peak, experimentally observed at $\omega =0$ in the dynamical
structure factor above the transition temperature, e.\,g.\ in
SrTiO$_3$ \cite{Ris71,Sha72}, can be explained as a consequence of
defects \cite{Hal76,Wag80a,Wag80,Schw91}. In recent years it has been
found that the critical fluctuations above \tc{} reveal a second
length scale in \srt{} as well as in many others systems
\cite{And86a,Rya86,Thu93}, a feature which has been reviewed by
R.~Cowley \cite{Cow96a}. This phenomenon was observed in surface-near
regions (typically $\approx 100$\,$\mu$m) \cite{Neu95b,Rue97a} and the
origin of the second length scale is considered to be the existence of
defects \cite{Wan98a}.\\ 
In the present paper, we concentrate on the
bulk behaviour where no sharp component occurs, also the central peak
is not subject of our investigations, the behaviour of the sharp
component will be discussed in a forthcoming paper (paper~II)
\cite{Hue00b}. We have probed the cubic-to-tetragonal phase transition
in SrTiO$_3$ at $T_c \approx 100$\,K with synchrotron radiation. The
high {\bf q}-space resolution and the high peak to background ratio
obtained with triple-axis diffractometry at HASYLAB allows for
detailed studies of the critical fluctuations at structural phase
transitions \cite{Neu94a,Rue95d}.\\ 
The critical scattering above
the antiferrodistortive transition of \srt{} has been studied for
samples with different oxygen vacancy concentrations. We measured the
order parameter and the critical exponents of the correlation length
and the susceptibility, the oxygen vacancy concentration in the
samples was determined independently. The experimental findings from
the scattering experiments are discussed quantitatively in terms of
the influence of defects. In the theoretical discussion the static
critical behaviour at structural phase transitions is considered with
a Ginzburg-Landau functional, taking into account the influence of
defects through a local increase of the critical temperature near the
defects. The use of the replica trick makes it possible to investigate
the static critical behaviour with the help of the
renormalization-group theory \cite{Edw75,Mez87}. However, in recent
years a possible break down of the symmetry of the replica functional
(replica-symmetry breaking RSB) is discussed
\cite{Mez87,Dot95a,Nar99,Mar99}. Until now the influence of the free
energy landscapes or ``rare regions'', which can occur for defect
systems, has not been studied for structural phase transitions. Here
the possibility of local order parameter clusters must be taken into
account.
 
The paper is organized as follows: In Sec.\,\ref{sec:exp} we present
the experimental details and the results of the scattering
experiments, followed by a description of the preparation and
characterization of the samples. Sec.\,\ref{sec:theo} provides the
theoretical analysis of the experimental data, assuming first the
qualitative behaviour and then the order parameter-states leading to
the minimization of the free energy within the partition
function. Finally, in Sec.\,\ref{sec:sum} we summarize and discuss our
results.
\section{Experimental results}
\label{sec:exp}
The scattering experiments have been performed on 5 different samples
using 120 and 100~keV synchrotron radiation at the undulator beamline
PETRA~II and the wiggler beamline BW5 at HASYLAB, respectively.  All
data have been collected at 3-axis-diffractometers in Laue geometry at
the position of the ($511$)/2 superlattice reflection. In order to get
both high incident photon flux and adequate {\bf q}-space resolution
in the scattering plane, annealed Si ($311$) crystals have been used
as monochromator and analyzer. The resulting resolution (HWHM) in the
scattering plane at BW5 (PETRA~II) was \Dq{x} =1.5--3\ang{-3}
(\Dq{x}=1\ang{-3}) in the longitudinal direction and 2\ang{-4} $\le$
\Dq{y} $\le$ 2\ang{-3} in the transverse direction, depending on the
mosaicity of the respective sample.  Perpendicular to the scattering
plane the resolution was of the order of \Dq{z}=1\ang{-1}.  The
experimental setup is described in detail in
Refs.\,\cite{Bou98,Rue97a}.  The deconvolution of the experimental
data was performed as described by Hirota et al.\,\cite{Hir95a}.\\ %
In order to study the effect of oxygen doping, bulk measurements have
been performed on a reduced, an oxidized and an as grown Verneuil
single crystal.  Additionally, a flux-grown sample \cite{Sche76} and
an almost perfect crystal, grown by means of the float-zone technique
\cite{Shi69}, have been investigated.  In Table\,\ref{tab:prep} the
different sample treatments and growth techniques are summarized.  The
behaviour of the sharp component in these samples will be discussed in
paper~II \cite{Hue00b}, in the following we only concentrate on the
bulk properties.\\ The mosaicity of the Verneuil-grown crystals varied
between 30'' and 100'' which is rather broad compared to the almost
perfect float-zone grown (1''-7'') and flux-grown samples (10'').
However, the latter two crystals are slightly brownish, perhaps due to
iron impurities \cite{Dar76}, in contrast to the transparent as-grown
Verneuil samples.  Impedance measurements have been performed in order
to determine the oxygen vacancy concentrations in the different
samples \cite{Den95a}.  Interestingly, the amount of oxygen vacancies
is about two orders of magnitude bigger for the crystallographic more
perfect samples compared to the as grown and oxidized Verneuil-grown
crystals.  However, the reduced sample shows significant changes: the
colour changes from transparent to black and the resistance decreases
substantially, i.\,e.\ the amount of oxygen defects increases about
three orders of magnitude.  Impedance measurements with
microelectrodes of different diameters \cite{Fle96,Rod99} revealed a 
spatially homogeneous conductivity in the samples I and II.
Owing to the relation between conductivity and oxygen vacancy concentration 
\cite{Den95a} it can be concluded that the concentration of the 
oxygen vacancies is distributed homogeneously over the respective samples.
Assuming a random distribution
of these vacancies the mean distance of defects results to
$d=n^{-1/3}$ with $n$ the defect concentration.  This distance varies
between $\approx 240$\,\AA{} for the lower defect concentrations and
$\approx 40$\,\AA{} for the higher concentrations [see Table
\ref{tab:samples}].\\ 
In Fig.\,\ref{fig:tc} the integrated intensity
of the superstructure reflection ($511$)/2, normalized to the
extrapolated value at $T=0$\,K, is plotted versus the reduced
temperature $\tau=\frac{T}{T_{\rm c}}-1$ for the different
Verneuil-grown samples and the float-zone grown sample.  Below \tc{}
the intensity is following a power law with the critical exponent
$2\beta=0.68$, which is well known for \srt{} \cite{Mue71,Ris71}.
Above \tc{} we observe a tail which is dependent on the defect
concentrations.  
The tail above the critical
temperature can originate from both the critical fluctuations and an
additional contribution to the squared order parameter caused by
defects.  This can be interpreted in the way that static order
parameter clusters exist at temperatures higher than \tc{}.
Nevertheless, for the determination of the critical temperatures we
followed the procedure explained in \cite{Ris71}, i.e.~at $T_c-\Delta
T$ half of the intensity at $T_c+\Delta T$ has been substracted from
the data of the integrated intensities of the superstructure
reflections, which corrects for the effect of the critical scattering
contribution to the integrated reflecting power of the superlattice
reflection.  The values of the critical temperatures, which have been
found for the different samples, are given in Table \ref{tab:samples}.
A shift of the critical temperature to lower values with raising
defect concentration is visible, which has been observed earlier in
Refs.\,\cite{Has78,Bae78}.  \\ 
The inverse correlation length
$\kappa_{\rm L}(T)$ and the susceptibility $\chi_{\rm L}(T)$ are
deduced from the critical scattering at the ($511$)/2 superlattice
reflection after deconvolution with the experimental resolution
function.  The critical scattering in the float-zone grown sample has
been measured over a much larger temperature interval than those of
the other samples.  Here, a crossover between different values for the
critical exponents $\nu$ and $\gamma$ obtained from $\kappa_{\rm
L}(T)$ and $\chi_{\rm L}(T)$ can be identified [see
Fig.\,\ref{fig:broad}].  These data have been taken in the bulk of the
float-zone grown sample but the probed volume also included two
surfaces (the front and the back side).  However, due to the
relatively large thickness (12\,mm) of the sample, the contribution of
the sharp component to the signal, originating from a thin layer at
the surface, was less than $1\%$ close to \tc{} and has been
neglected.  \\ Around $\tau_c=0.11$ a change of slope is clearly
visible for both $\kappa_{\rm L}(T)$ and $\chi_{\rm L}(T)$.  This
crossover temperature $\tau_c$ is in excellent agreement with the
value of $\tau_d$ [see Table \ref{tab:samples}], which is defined as
following: $\tau_d$ is the reduced temperature at which the
correlation length $\kappa^{-1}_L(\tau_d)$ is equal to the mean
distance $d$ of the oxygen vacancies.  Above this crossover
temperature the system can be described within mean-field theory,
which leads to exponents $\nu=0.5$ and $\gamma=1.0$, but below the
crossover temperature the fitted exponents $\nu=1.19(4)$ and
$\gamma=2.89(4)$ are much higher than theoretical values.  The
non-classical exponents expected for this 3d-Heisenberg system with
cubic symmetry, calculated by LeGuillou et al.\,\cite{LeG80}, are
$\nu=0.705(3)$ and $\gamma=1.386(4)$.\\ 
In Figure \ref{fig:all_broad}
the inverse correlation length for the four other samples is plotted
against the reduced temperature.  Effective exponents $\nu$ and
$\gamma$ have been derived by fits of power laws to each data set
without taking into account the crossover scenario, because the
expected crossover temperature $\tau_d$ is either too small ($\approx
0.02$) or too large ($0.20$) to allow an observation of the crossover
within the experimental probed temperature range.  The effective
exponents vary between $\nu\approx 0.7$, $\gamma\approx 1.4$ (as grown
and oxidized Verneuil-grown samples) and $\nu\approx 1.2$,
$\gamma\approx 2.4$ (reduced Verneuil-grown sample).  It should be
emphasized that the scaling relation $\gamma=(2-\eta)\nu$, with
$\eta\approx 0.03$ \cite{Fis64}, still holds!  Also, the scaling
relation $2\beta=d\nu-\gamma$, which yields $\beta=0.34(6)$ for the
float-zone grown sample, is in excellent agreement with the
experimental data below \tc{}.\\ In the following discussion an
attempt is made to understand the microscopic nature of these
experimental results.  %
\section{Theoretical discussion}
\label{sec:theo}
\begin{figure}[ht]
  \begin{center}
    \includegraphics[angle=-90.0,width=\linewidth]{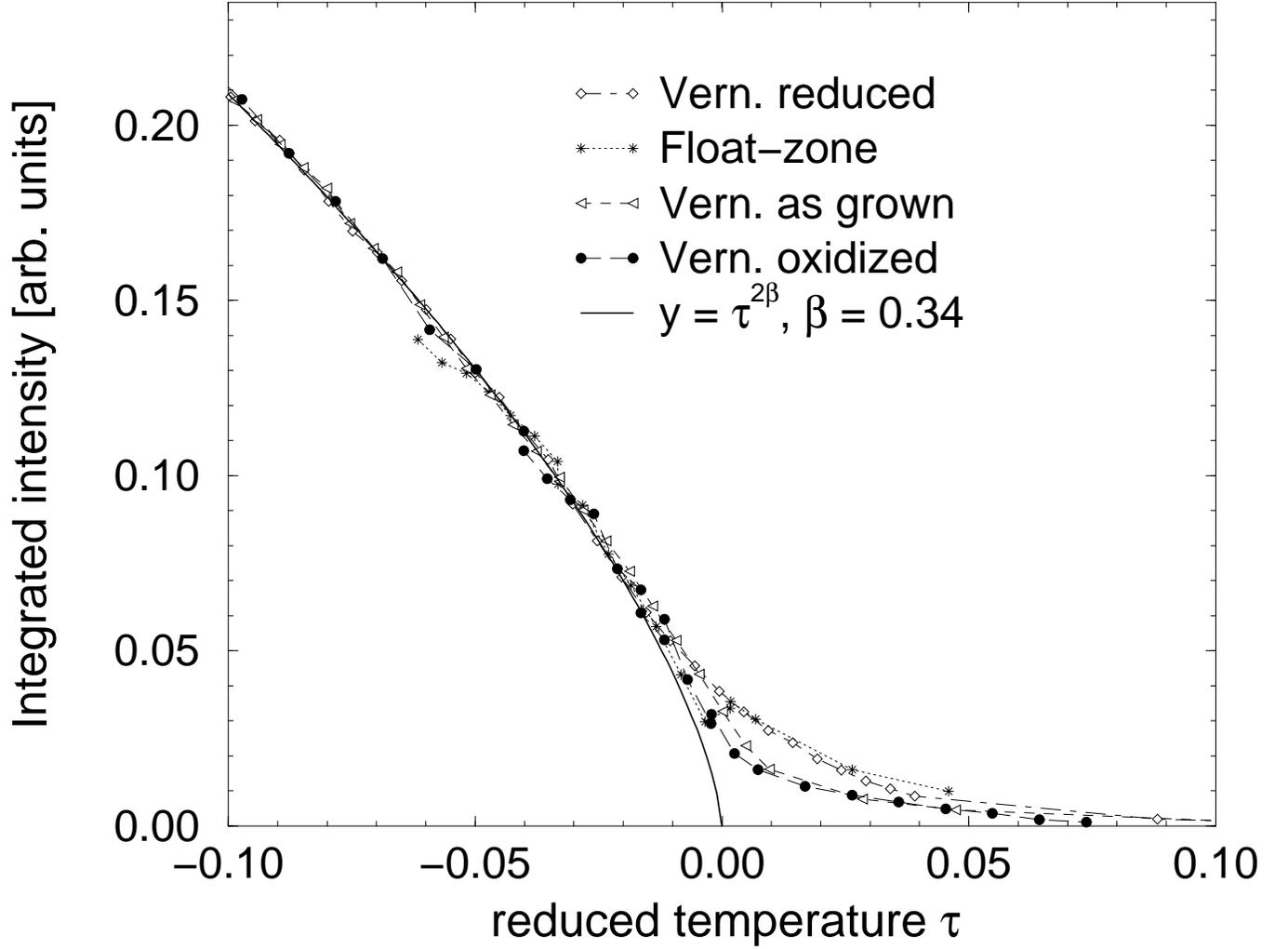}
  \end{center}
 \caption{Integrated intensities of the superlattice reflection ($511$)/2. In order to compare the tails above \tc{} the intensities have been normalized such that they coincide below the critical temperature. The critical exponent $\beta$ has been fixed to $\beta=0.34$. \label{fig:tc}}
\end{figure}

In our experimental measurements we found a smearing-out of the
squared order parameter near the phase transition, which can be
explained theoretically by the presence of local order parameter
clusters above \tc{} \cite{Hal76,Sch78,Schw91}.  Defects, which
locally increase the transition temperature, lead to order parameter
clusters above the critical temperature of the homogeneous system.
For SrTiO$_3$ we expect a rotation of the oxygen octahedra near the
position of oxygen vacancies to be responsible for finite order
parameter clusters above \tc \cite{Sat85}.  These rotations can occur
around one of the three principal directions.  Thus a three
dimensional Heisenberg model should be used to model the critical
behaviour.  In Fig.\,\ref{fig:tc} the solid line shows the fit to the
experimental data using the power law for the normalized integrated
intensity versus the reduced temperature I$^{1/2}\propto \tau^{0.34}$.
Because $I^{1/2}$ is proportional to the order parameter, the value
for the critical exponent $\beta$ is in good agreement with the values
obtained for the three dimensional Heisenberg model in the homogeneous
case \cite{LeG80}.  But, in clear difference to the critical behaviour
of homogeneous systems is the fact that for higher concentrations of
oxygen vacancies a more pronounced tail exists [see Fig.\,\ref{fig:tc}
and Table \ref{tab:samples}].  Of course the critical fluctuations
also contribute to the strength of these tails.  However, the
contribution of the existing order parameter clusters and the
dependence of the defect concentration is obvious.  Unfortunately, for
the experimental data investigated here, these two contributions
cannot be differentiated adequately.  Nevertheless, we clearly
identified the first clear effect of defects on the critical
behaviour.  In an additional remark, we want to stress that our
observation of a finite value of the squared order parameter does not
mean that there exists a finite bulk value for the order parameter
above \tc.  The local order parameter clusters can be oriented in
different directions in the order parameter space.  If all clusters
are oriented in one direction, a finite bulk value for the order
parameter would result [see Ref. \cite{Schw91}].

\begin{figure}[ht]
  \begin{center}

    \includegraphics[angle=-90.0,width=\linewidth]{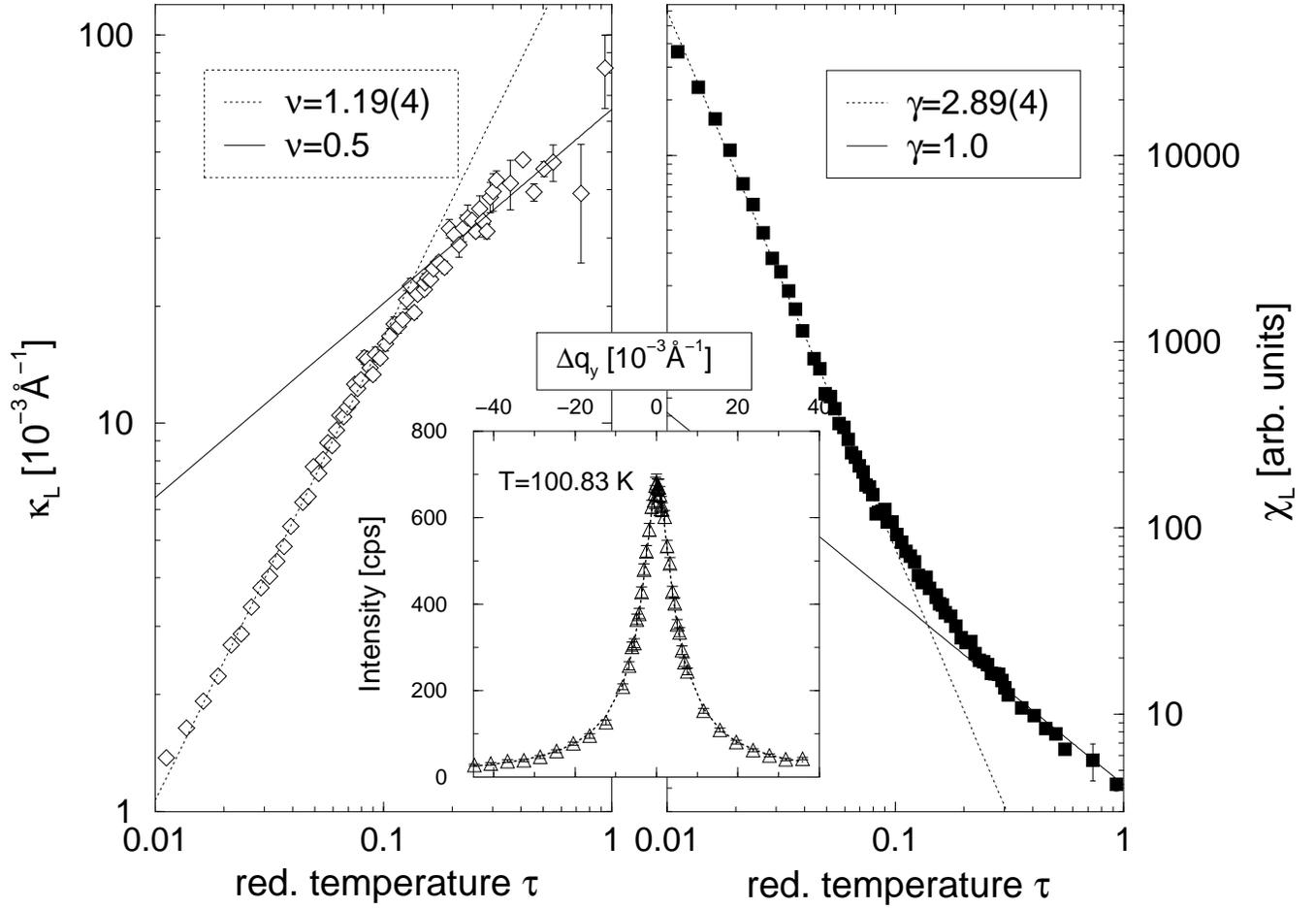}
  \end{center}
 \caption{Inverse correlation length and susceptibility of the broad component in the bulk of the Float-zone grown sample. Around $\tau_c = 0.11$ a crossover in the critical behaviour is clearly visible. The inset shows a transverse scan and the best fit to the data at $T=100.8$\,K. \label{fig:broad}}
\end{figure}

\begin{figure}[ht]
  \begin{center}
    \includegraphics[angle=-90,width=\linewidth]{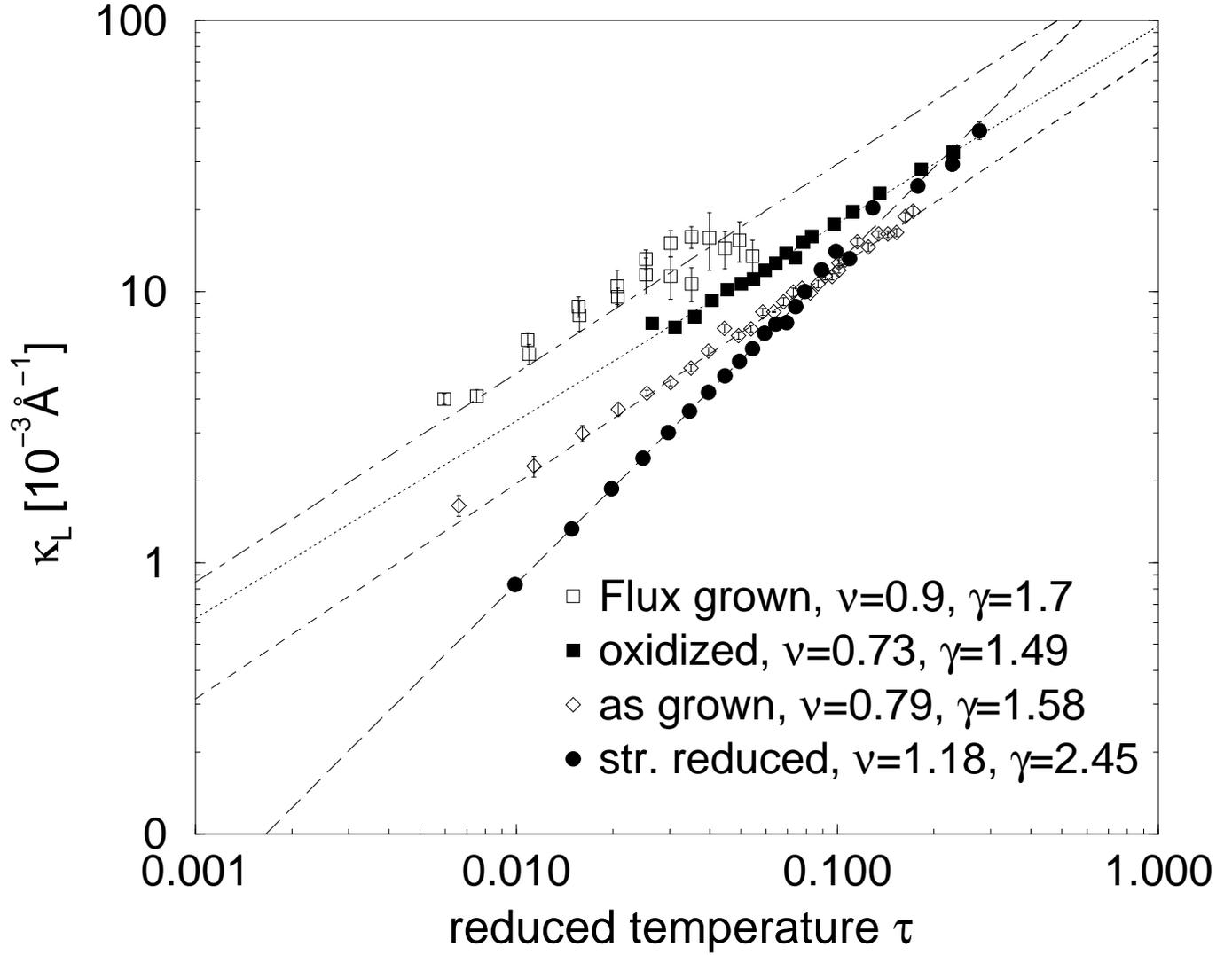}
  \end{center}
 \caption{Inverse correlation length for different samples. The effective exponents $\nu$ and $\gamma$ have been derived from power law fits to the inverse correlation length and the susceptibility. \label{fig:all_broad}}
\end{figure}

For the float-zone grown crystal a crossover 
of  the critical exponents $\nu$ and $\gamma$ can be 
clearly observed as another effect of the defects
[see previous Section], taking place at $\tau \approx 0.11$
[see Fig.\,\ref{fig:broad}].
This is the temperature range where the squared order parameter 
starts to be noticeable different from zero 
[see Fig.\,\ref{fig:tc}].
In the general picture of the influence of defects 
on the critical behaviour these findings can be understood as follows.
Far away from the critical region  mean-field
exponents are expected, because fluctuations are small
\cite{Lev59,Gin60}.
These mean-field exponents are found indeed for the float-zone 
grown crystal for $\tau>0.11$  [see Fig.\,\ref{fig:broad}]. 
Lowering the temperature ($\tau \le 0.11$) one has to take into
account fluctuations and a crossover to non classical critical 
exponents should occur.  
Without defects, one would expect the critical exponents for the 
3D Heisenberg model (e.\,g.\ $\nu \approx 0.705$ \cite{LeG80}).
We find near the critical temperature for the float-zone grown 
crystal the value of the exponent $\nu = 1.19$ 
[see Fig.\,\ref{fig:broad}], which differs
substantially from the one stated above.
Because the correlation length becomes larger near the critical 
temperature, the critical fluctuations can become correlated over distances 
larger than the mean distance between the defects.
In this case the impurities can change the critical behaviour and
a crossover to a different critical exponent is possible.
The Harris criterium defines the necessary condition \cite{Har74a}. 
If this condition is fulfilled, a dependence of the crossover 
temperature from the defect concentration should be observable. 
For the crystals probed here, with their different defect
concentrations,  we, indeed, find such a dependence 
[see Figs.\,\ref{fig:broad},\ref{fig:all_broad} and 
Table \ref{tab:samples}].
As described above the float-zone grown crystal has a crossover 
temperature equal to $\tau_d=0.11$.
For all other samples, except the reduced crystal, the characteristic 
temperature $\tau_d$ is very small ($\le 0.023$) and thus the 
crossover itself is not visible in the experimental data.
However, the critical exponents, obtained from power law fits to 
the data-sets in Figure \,\ref{fig:all_broad}, are smaller than 
$\nu=1.19$ but larger than the Heisenberg-value $0.705$.
The crossover from the Gaussian exponents (e.\,g.\ $\nu=0.5$) at 
high temperatures to the ``defect-induced'' exponents 
(e.\,g.\ $\nu\approx 1.2$) close to the critical temperature 
should lead to effective exponents $\nu_{eff}$ with values in the
range $0.5<\nu_{eff}<1.2$, as observed.

For the strongly reduced sample the crossover takes place 
at very high temperatures ($\tau>0.2$) so that the exponent $\nu=1.2$,
similar to the value observed in the float-zone grown crystal, 
is found in the whole temperature range probed here.

This again is in good agreement with the assumption that 
localized defects are responsible for the change in 
the critical behaviour.
In the regime near the critical temperature \tc{} a critical 
behaviour different from the homogenous case, i.\,e.\ other critical 
exponents, should be observable. 
The critical exponents for systems with quenched disorder 
have been calculated before \cite{Har74a,Khm75,Gri76}.
Those values can be compared with the critical exponents found 
in our experiments (e.g. in an $\varepsilon$-expansion 
for a 3-component order parameter
$\nu = 0.5 + \varepsilon \cdot 9/64 + ... \approx 0.64$ with 
$\varepsilon = 4 - D =1$ in three dimensions \cite{Gri76}).
Unfortunately, we find in our experiments  much higher values 
for $\nu$ and $\gamma$.
One important question arises immediately: can we understand these findings 
within the concept of universality or are we facing non-universal
behaviour? 
A first hint is that we find the 
scaling relation $\gamma = (2-\eta) \nu$ still valid for all values of the 
critical exponents for the different crystals near the 
critical temperature [see Figs.\,\ref{fig:broad},\ref{fig:all_broad}].
As already discussed in the last section, the scaling relation 
for $\beta$ holds, too.

In order to fully understand the experimental data in terms
of the influence of defects on the critical behaviour the high values 
for the critical exponents $\nu$ and $\gamma$ near \tc{} need to be
explained. 
In the next two subsections we try to provide a theory which can 
explain the observed values for the exponents and 
can model the crossover behaviour.

\subsection{Defect-functional}
\label{def_func}
To investigate the questions arising from our experimental findings, 
the influence of local order parameter clusters, occuring 
above the critical temperature \tc{}, on the critical behaviour 
is taken into account.
At low defect concentrations the 
order parameter clusters are supposed to be well isolated.
Therefore, in the mean-field picture 
multiple local minima solutions exist,
because the localized clusters can have different orientations in 
order parameter space \cite{Sch78}. 
The properties of such ground states were discussed for spin-glasses 
\cite{Mez87}, random-ferromagnets \cite{Dot95a}, and recently for 
quantum magnets \cite{Nar99}.  
For random-ferromagnets Dotsenko et al.\,\cite{Dot95a,Dot95b}
found that the exponents calculated with the help of the replica-method 
are unstable with respect to a possible breaking of the replica
symmetry.  
New exponents result, different from those obtained by the usual
replica-method. 
Even scenarios for possible instabilities of the theory are given.
From these studies it is obvious that the states of minimal free
energy in the partition function have to be considered in detail. 
In the present paper we sketch the theoretical derivation,
for a more detailed discussion see e.g. \cite{Dot95a,Dot95b,Kau00}.

The starting point for the theoretical discussion is 
a Ginzburg-Landau model for continuous structural phase transitions in
disordered systems with quenched impurities. 
Until recently, in the standard approach only fluctuations around 
the homogeneous ground state of the order parameter
$\varphi(x) =0$ were considered in the replica-method  \cite{Dot95a}.
We will incorporate fluctuations around the states with the lowest free 
energy in the partition function.
The defect-functional used for the investigation of the effects of 
impurities on the free energy \cite{Schw91} reads  
\begin{equation} 
\label{def_funk_theo}
{\cal F} [ \varphi ({\vc x}); U  ] = \int_{}^{} d^dx 
\left[ 
\frac{1}{2} \, \sum_{i=1}^{k} \left(   
( a + U ) ( \varphi^{i}({\vc x}))^2 
+ (\nabla\varphi^i({\vc x}))^{2} 
\right)
+  \frac{b}{4} \, [\sum_{i=1}^{k}(\varphi^i({\vc x}))^2]^2 \right] 
\end{equation} 
Here $i=1\dots k $ denotes the component of the order parameter field,
$a=a'(T-T_c^0)$ is the harmonic coefficient and vanishes at the bulk transition
temperature $T_c^0$ of the pure system.
The coefficient $b$ of the non-linear term has to be positive.
With the short-range potentials 
\begin{equation}
U = \sum_{i_{\rm D}=1}^{N_{\rm D}}
                   U({\bf x}-{\bf x}_{i_{\rm D}}) \ .
\end{equation}
the effects of randomly distributed defects at the
sites ${\bf x}_{i_{\rm D}}$ ($i=1 \dots N_{\rm D}$) are described. 
They can be considered as point like defects. 

Applying the replica-trick \cite{Mez87}, the averaging over different
defect-distributions can be carried out, only taking into account fluctuations
around the homogeneous ground state $\varphi({\vc x})=0$. 
The discussion of the resulting fixed-points and flow equations can
be found in Refs. \cite{Gri76,Kor96}.

\begin{figure}[ht]
  \begin{center}
    \includegraphics[width=\linewidth]{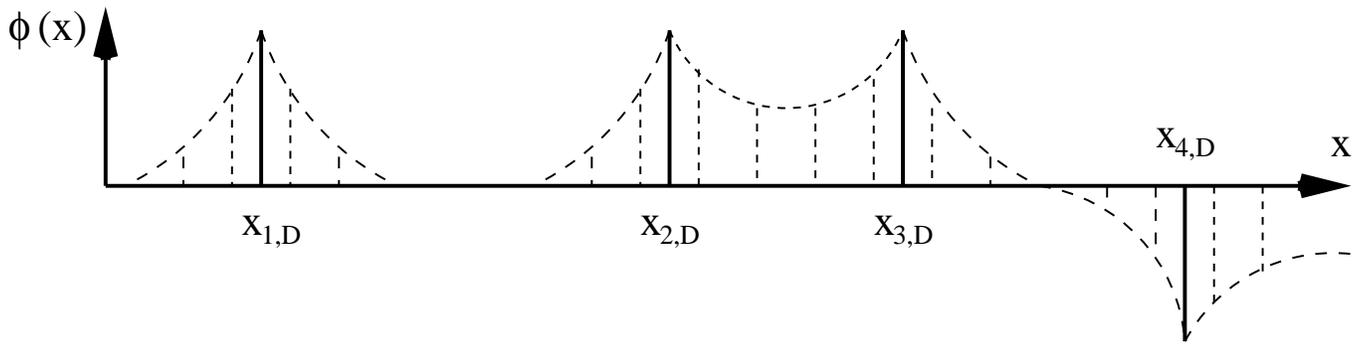}
  \end{center}
 \caption{A typical order parameter configuration above the critical
   temperature is sketched.  Thus, in the presence of localized order
   parameter clusters a complicated solution of the ground state
   results. \label{fig:op_clus}}
\end{figure}

To study the additional effect of possible order parameter 
clusters we start with the saddle-point approximation  
\begin{equation}
-  \Delta\varphi({\vc x}) + ( a + \sum_{i_{\rm D}=1}^{N_{\rm D}}
                   U({\bf x}-{\bf x}_{i_{\rm D}}) ) \varphi({\vc x}) 
 + b \, \varphi({\vc x})^3 = 0 \ ,
\end{equation}
which gives the minimum solutions of the free energy.
To keep the discussion transparent we choose the number of order-parameter
components equal to one ($k=1$).
As far as we consider defects which locally increase the
transition temperature, local order parameter 
condensates exist above \tc, as demonstrated in 
earlier studies \cite{Sch78,Schw91}.
In Fig.\,\ref{fig:op_clus} a typical
order parameter configuration is sketched.
Thus in the presence of localized order parameter clusters, 
a more complicated solution of the ground state is found. 
However, as long as the concentration of the defects and the 
order parameter clusters is small, the value of $\varphi(x)$ outside
these regions decays exponentially.
The tail of such localized order parameter-clusters can approximately be
written as $\varphi_{lok;0}^{i} ({\vc x}) \propto \frac{1}{|{\vc x} 
- {\vc x}_{i_{\rm D}}|} 
\, \e^{-|{\vc x} - {\vc x}_{i_{\rm D}}| \, / \ell}$ \ \cite{Sch78,Schw91}.
Therefore, an approximation of the extremal solution is
\cite{Dot95a,Nar99}
\begin{equation}
\Phi^{(p)} ({\vc x}; U) = \sum_{i=1}^{N_{\rm D}}
\sigma_i^{(p)} \cdot \varphi_{lok;0}^{i} ({\vc x}) \ ,
\end{equation}
with $p=1 \dots L = 2^N$ possible solutions because of the 
Ising symmetry in the order parameter space ($k=1$) for a given
defect distribution $U$ and $\sigma_i^{(p)}=\pm 1$.
No finite order parameter results for well isolated and
non-interacting order parameter clusters \cite{Schw91}.
The clusters are of course temperature dependent, but we 
neglect this dependence here, assuming that in the temperature range
where their contribution is most important the cluster structure 
does not change that much. 
Now we can write the partition function $Z$ in a way that we 
consider the global minimum solutions $\Phi^{(p)}$ and 
fluctuations $\varphi(x)$ around these solutions
\begin{equation}
Z[U] =  \int D\varphi({\vc x}) \sum_{p=1}^{L}
\e^{- {\cal F} [
\Phi^{(p)} [ {\vc x}; U ] + \varphi({\vc x});U ] } 
=  \int D\varphi({\vc x}) \e^{- {\cal F} [\varphi;U ] }
\cdot \tilde{Z}[U] \ ,
\end{equation}
where the part of $Z$ with the contributions from the saddle point solutions
is written as
\begin{equation}
\tilde{Z}[U] = \sum_{p=1}^{L}
\e^{- {\cal F} [\Phi^{(p)};U] } \cdot \e^{ -\int d^{d}x [
\frac{3 \, b}{2} \, \left( \Phi^{(p )} 
[{\vc x};U] \right)^2 \varphi({\vc x})^2 +
b \, \Phi^{(p)} [{\vc x};U] \varphi({\vc x})^3 ] } \ . 
\end{equation}
$Z$ and $\tilde Z$ depend on the actual defect potential $U$,
because the free energy ${\cal F}$ does.
Thus, we are still investigating the effect of  frozen disorder. 
Because the system with defects is no longer translationally invariant,
the trace in the partition function can no longer be treated 
in the usual way.
The replica-method will be used to deal with the problem.

An additional remark concerning the energy barriers should be made.
In our approach the existence of large energy barriers between the
various saddle-point configurations is assumed. Unlike in glasses
these barriers are not infinite.
But as already pointed out by Narayanan et al.\,\cite{Nar99} in the
thermodynamical limit these barriers tend to infinity.
As a consequence also the dynamics of the clusters should be very
slow.
This gives us the possibility to consider just the statics and to
neglect effects of relaxation and reorientation.

Performing the disorder average with the help of the
replica trick by considering $n$ replicas, 
the effect of order parameter clusters can be incorporated. 
With the saddle-point solution we find
\begin{equation}
\label{sadpoi}
Z_{n} = \langle \int D \varphi_1 \dots D \varphi_n \ \e^{ 
- \sum_{\alpha=1}^{n} {\cal F } [\varphi_\alpha;U]}
\cdot \tilde{Z_{n}}[\varphi_1 \dots \varphi_n; U] \, \rangle_U \ ,
\end{equation}
with the replica index $\alpha$ and
\begin{equation}
\tilde{Z_{n}}[\varphi_1 \dots \varphi_n; U] = 
\sum_{p_{1}...p_{n}=1}^{L} 
\e^{-\sum_{\alpha=1}^{n} {\cal F} [\Phi^{( p_\alpha )};U]} 
\cdot 
\e^{- \int d^{d}x \sum_{\alpha=1}^{n}
[ \frac{3 \, b}{2}  \, \Phi^{(p_\alpha)} 
({\vc x})^2 \varphi_\alpha({\vc x})^2 +
b \, \Phi^{(p_\alpha)} ({\vc x}) \varphi_\alpha({\vc x})^3 ] } \ .
\end{equation}
This is the so called replica partition function for our problem.
In the following the additional effect of order parameter clusters 
should be studied, thus the contribution of $\tilde{Z_{n}}[U]$ 
is our objective.
The treatment of this kind of replica partition function was studied
by Dotsenko et al. \cite{Dot95a,Dot95b}. 
The authors analyzed the problems arising from this complicated 
structure for the solution of the partition function.
As a result, they introduced two approximations to allow a 
treatment within the usual renormalization group theory, which will 
be discussed in the following.

Firstly, a distribution function for the local order parameter clusters  
$\varphi_{lok;0}^{i} ({\vc x})$ is presumed \cite{Dot95a}.
With that we interpret the solutions arising from the saddle-point
approximation as random variables,
even though for a given distribution of defects these solutions 
are fixed.
However, we want to sum over all possible different distributions,
hence this treatment is adequate.
A Gaussian distribution function \cite{Dot95a}
\begin{eqnarray}
P[ \{ \varphi_{lok}({\vc x}) \} ]  \equiv P[ \Phi^{(p)} ({\vc x};U)]
&=& \prod_{i=1}^{N_{\rm D}} \e^{ 
-\frac{1}{2\Delta}\int d^d x [\varphi_{lok}^{i}({\vc x})^2 
- \varphi_{lok;0}^i({\vc x})^2 ]^2 }  \nonumber \\
 &=& \e^{- \frac{1}{2\Delta} \int d^d x
[ \Phi^{(p)} ({\vc x})^2 -\Phi_0 ({\vc x})^2 ]^2} \ ,
\end{eqnarray}
with $\varphi_{lok}^{i}({\vc x})$ as the local order parameter cluster
is used. 
The parameter $\Delta$ depends on the distribution of the
defect-strength, the temperature,  and the coupling constant $b$.
Furthermore, $\Phi_0({\vc x})$ is the global solution $\Phi^{(p)}$ with 
$\varphi_{lok}^{i}({\vc x}) = \varphi_{lok;0}^{i} ({\vc x})$.
In this approximation $\Delta$ is treated as a constant 
parameter.
The crucial point here is the assumption that $\Phi^{(p)}$ 
still is a solution of equation (\ref{sadpoi}).
Otherwise new interaction terms would be introduced in the
functional.

Now the fluctuations have to be considered. 
For $\Delta$ small, 
$\Phi^{(p)}({\vc x})^2 = \Phi_0({\vc x})^2 + \delta \Phi^{(p)} ({\vc x})$,
the fluctuations $\delta \Phi^{(p)} ({\vc x})$ around the 
saddle point solutions can be integrated out and 
it results \cite{Dot95a}
\begin{eqnarray}
\tilde{Z_{n}} &\simeq&  \prod_{p=1}^{L} 
\Biggl[ \int D \, \delta \Phi^{(p)}({\vc x})
\e^{ -\frac{1}{2\Delta}
\int d^{d}x \ \delta \Phi^{(p)} ({\vc x})^{2} } \Biggr] \\
 && \times\sum_{p_{1}...p_{n}=1}^{L} \e^{ \
\frac{b}{4} \int d^d x \, \sum_{\alpha=1}^{n} 
\delta \Phi^{(p_{\alpha})} ({\vc x})^2 
- \frac{b}{4} \int d^d x \, \sum_{\alpha=1}^{n} 
\delta \Phi^{(p_{\alpha})}({\vc x})
[6\varphi_{\alpha}({\vc x})^2 - 2 \Phi_{0}({\vc x})^2 ]} \ , \nonumber
\end{eqnarray}
defining $\Delta' = (1/\Delta - b /2)^{-1}$.

As a second approximation Dotsenko et al. suggested a heuristic ansatz
to derive the leading contribution to the partition function $\tilde
Z_n$ \cite{Dot95a}.
From the solution of the random-energy-models (REM) 
it is known that a partition function like $\tilde Z_n$ takes 
its maximum for $1 \leq x_{0} \leq n$. 
For the REM it's possible to calculate the parameter $x_0$ by
extremizing the resulting free energy with respect to $x_0$.
Unfortunately, in our problem the integration over fluctuations is done in
a perturbative approach. 
This is why we cannot calculate the value of $x_0$ for the problem
at hand explicitly.
For a detailed discussion see Ref. \cite{Dot95a}.

Applying the two approximations stated above, a breaking of the 
symmetry of the replica-functional turns out to be
significant for the Ginzburg-Landau-Wilson functional discussed here 
\cite{Dot95a,Kau00}, which is a consequence of the fluctuations 
around the minimal free energy solutions $\pm\Phi_0 ({\vc x})$. 
The form of the resulting functional
was first derived by Dotsenko et al. \cite{Dot95a}.
Because the resulting form of the functional considered here is
equivalent to the one discussed in Ref. \cite{Dot95a}, we want 
to refer for all the technical details to Ref. \cite{Dot95a} 
and Ref. \cite{Dot95b}, especially for  
the derivation of the renormalization group flows and the calculation
of the fixed points. 
To keep our discussion tight, we will focus on the interpretation
of the results.

\subsection{Crossover scenarios}

After having shown how the breaking of the replica-symmetry can
occur for the defect functional in Eq. (\ref{def_funk_theo}), 
now the consequences for the critical exponents are discussed.
To analyze the crossover effects we concentrate on 
the effective exponent $\nu_{eff}$ of the correlation length
as a function of the reduced temperature $\tau$.

From the renormalization group flow eqations, which are calculated 
to one-loop order \cite{Dot95a}, the Gaussian fixed-point and 
the Heisenberg fixed-point occur. They both prove to be unstable.
But with the number of order parameter components $k$ in $1<k<4$ 
a third, stable fixed point is possible.
For example, the result for the critical exponent $\nu$ is
\begin{equation}
\nu = \frac{1}{2} + \varepsilon \, \frac{1}{2} \, 
\frac{3 k (1-x_0)}{16 (k-1) - k x_0 (k+8)} \  \label{nu_rsb} \ ;
\end{equation}
with $x_0 = 0$ the usual replica symmetric value of the 
critical exponent $\nu$ is obtained \cite{Khm75}.
Hence the effect of the breaking of the replica symmetry
(where $x_0 \neq 0$) on the 
critical exponents can be interpreted as a reduction of the
number of order-parameter components in a subtle manner.

\begin{figure}[ht]
  \begin{center}
    \includegraphics[angle=-90,width=\linewidth]{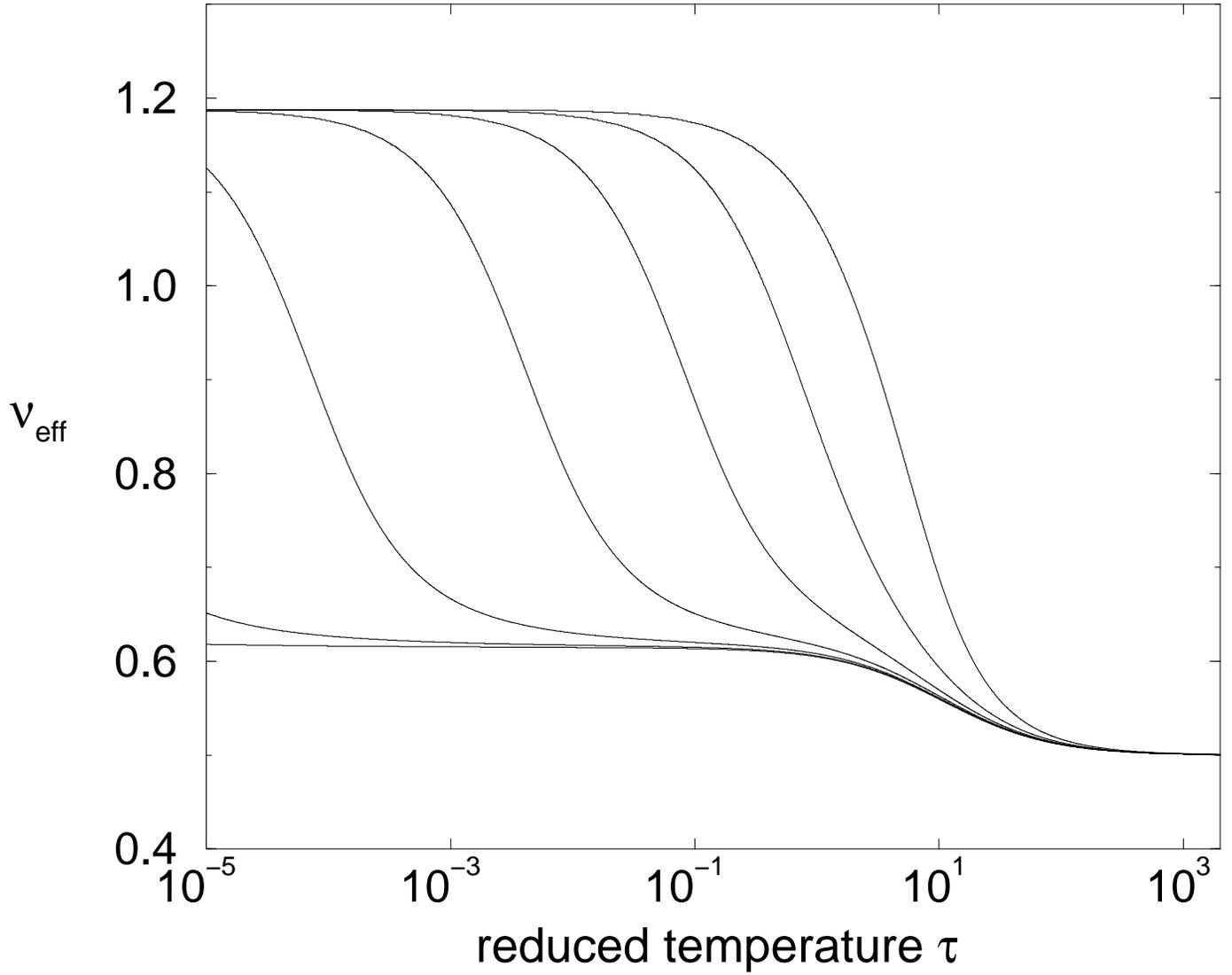}
  \end{center}
 \caption{The effective exponent $\nu$ 
   is plotted against the temperature-scale for different initial
   values of the coupling constants. $x_0$ is choosen as $0.9622$. \label{fig:nu_flow}}
\end{figure}

\begin{figure}[ht]
  \begin{center}

    \includegraphics[angle=-90,width=\linewidth]{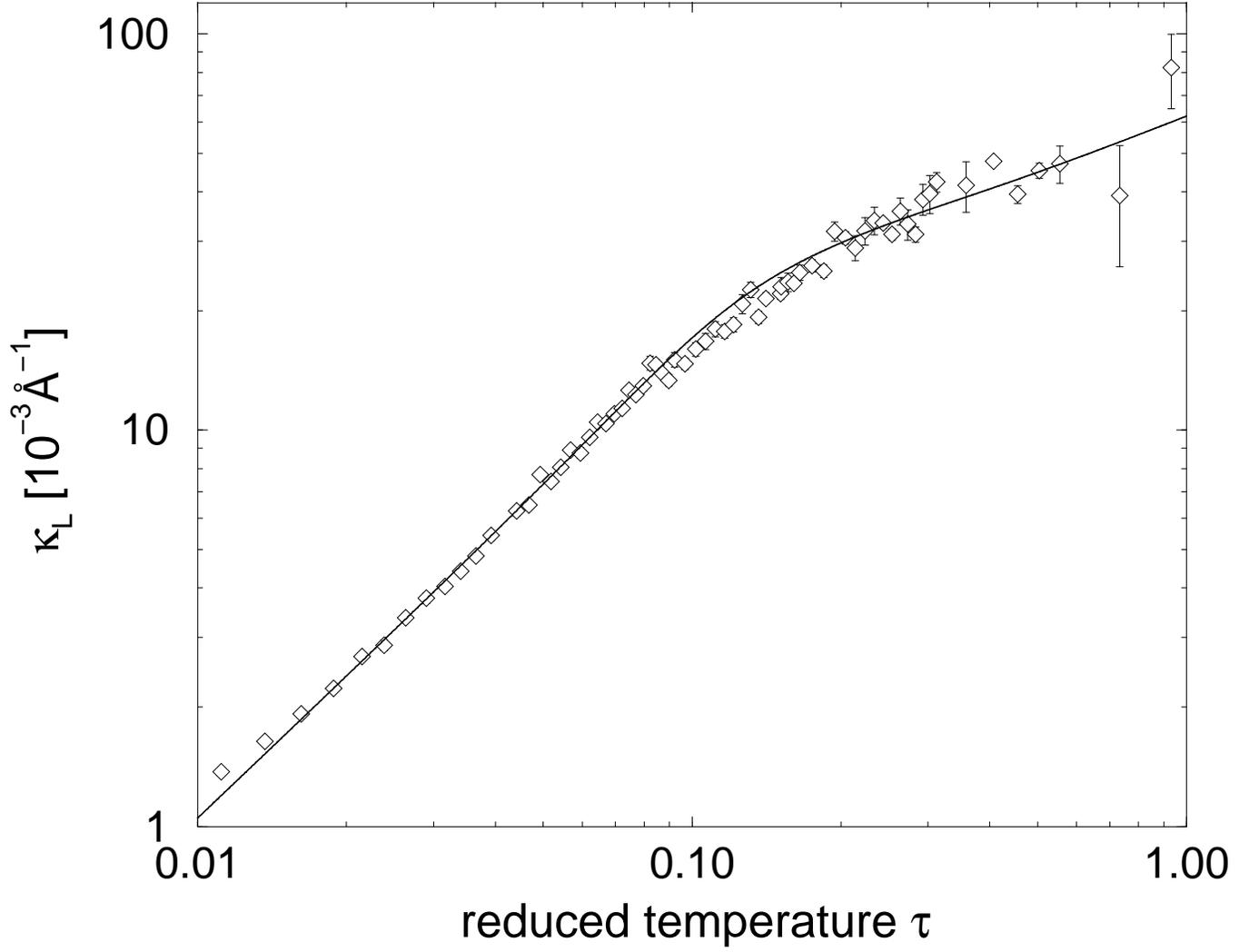}
  \end{center}
 \caption{Inverse correlation length of the broad component in the bulk of the
   float-zone grown sample. The theoretical curve models the
   crossover. \label{fig:kappa_fit}}
\end{figure}

As mentioned above, the value of $x_0$ cannot be derived in the
approximation considered here.
The experimental finding that asymptotically near \tc{} the values of the
critical exponent $\nu$ are very similar, hints to the fact that there
actually is only one $x_0$ for a particular class of defect systems.
In Fig.\,\ref{fig:nu_flow} a set of possible flows of effective 
exponent $\nu_{eff}$ versus the reduced temperature is sketched. 
Depending on the initial values for the coupling constant 
$b$ and the effective strength of the defects $c$ \cite{Dot95a,Kau00},
different crossover scenarios can result.
Starting from the Gaussian fixed-point (large $\tau$) 
the effective exponent $\nu_{eff}$ starts with its mean-field value.
For smaller $\tau$ the Heisenberg fixed-point, 
calculated to one-loop order ($1/\nu_{eff} = 2 - \frac{k+2}{k+8}$), 
is reached, if the initial value for $c$ is small.
For larger initial values of the defect strength  
a direct crossover from the mean-field fixed-point
to the ``defect-induced'' fixed-point is found.
The width of the temperature where the crossover occurs, also
varies with the initial values of $c$ and $b$.
Assuming $x_0=0.9622$, we can obtain the experimentally found 
critical exponent $\nu_{eff}=1.19$ for $\tau \rightarrow 0$, 
that is in the critical regime.
The temperature scale where the crossover takes place 
is dependend on the initial values of the coupling constants. 
By matching the temperature scale \cite{Kau99}, 
the crossover can be modeled, too.
For the float-zone grown sample the resulting fit of the 
experimental data is shown in Fig.\,\ref{fig:kappa_fit}.

Despite the remarkable agreement between theory and experimental
findings one should be aware of the fact that this is only a one loop
result. 
With a higher order in a loop expansion a similar fit might well
require a different value for $x_0$.
Through the two approximations, defined in the last
subsection, a lot of physical effects which are very hard to 
model are condensed into the parameter $x_0$.
The next goal of the theory should be to determine the value of
$x_0$ and to answer the question if this parameter depends 
on the defect strength or the defect concentration.
Dotsenko et al. expressed their hope that $x_0$ may be determined in
higher order in $\varepsilon$, in which case universal exponents for a
paticular class of defect sysems would result.
\section{Discussion}
\label{sec:sum}
In summary, we have investigated the effects of defects on the
critical behaviour at structural phase transitions.  Our experiments
probed the $105\,K$ phase transition of \srt{} and the influence of
oxygen vacancies for different \srt{} crystals.

We found in our experiments that the squared order parameter is
non-vanishing and rounded above the critical temperature \tc{},
representing the transtion temperature for a perfect crystal.  The
rounded squared order parameter increases above this temperature with
increasing concentration of oxygen vacancies in the crystals.

These effects were interpreted as a result of local order parameter
clusters occuring above \tc{} (see Ref. \cite{Schw91}).

Furthermore, near \tc{} unusual high values for the critical exponents
$\nu$ und $\gamma$ were found in our experiments.  For the float zone
grown crystal we were also able to study a crossover of the critical
exponents $\nu$ and $\gamma$ in great detail.  In addition, we
identified a dependence of the crossover-temperature on the defect
concentration of the respective samples.

The crossover scenario was analyzed in the framework of the
renormalization-group theory (see also Refs. \cite{Tae92,Kau99}). Two
main effects could be studied here.  Firstly, the correlation length
of the critical fluctuations has to exceed the length scale of the
mean distance between oxygen vacancies in order to change the values
of the critical exponents of the homogeneous system.  Secondly, the
unusual high values for the critical exponents found experimentally
below the crossover temperature, which indicate the crossover from
classical to non-classical critical behaviour, could be explained by
taking into account the fluctuations around the states of minimal free
energy and the resulting breaking of the replica symmetry.

Because of the occurence of the paramter $x_0$ within the theory,
which cannot be determined explicitly, some questions concerning the
universal aspects of the results are still open.  Two scenarios are
possible: (i) the parameter $x_0$ is universal for certain classes of
defect systems; (ii) the parameter $x_0$ depends on microscopic
properties of the crystals and is not universal.  We stated some hints
that for the class of defect systems studied here universal behaviour
occurs.

It is known from defect models \cite{Schw91} that in three dimensions
there is a minimum strength of defects. Below this border no order
parameter clusters form and the analyzed crossover should not occur.

It would be interesting to study similar systems in the presence of
different kinds of defects or to study systems where the universality
class can be changed, e.g. under pressure. Also further theoretical
investigations are necessary to get a step beyond the approximations
considered in this and in other papers that take into account the
breaking of the replica-symmetry.
\section*{Acknowledgements}
We would like to thank S.~Rodewald and J.~Fleig from MPI Stuttgart for
the conductivity measurements, G.~Shirane and H.J.~Scheel for making
available the float-zone and flux-grown SrTiO$_3$ crystals as well as
S.~Kapphan for preparing the Verneuil crystals, and U.~C.~T\"auber and
E.~Courtens for fruitful discussion.  Support from the DFG under
contract number Schw12-1 is acknowledged.
\newpage
\bibliographystyle{prsty}

\begin{thebibliography}{10}
  
\bibitem{Har74a} A. Harris, J. Phys. C {\bf 7}, 1671 (1974).
  
\bibitem{Khm75} D. Khmel{'n}itski{\u i}, Sov. Phys.-JETP {\bf 41}, 981
  (1975).
  
\bibitem{Gri76} G. Grinstein and A. Luther, Phys. Rev. B {\bf 13},
  1329 (1976).
  
\bibitem{Hal76} B. Halperin and C. Varma, Phys. Rev. B. {\bf 14}, 4030
  (1976).
  
\bibitem{Str80} B. Strukov, J. Taraskin, K. Minaeva, and V. Fedorikh,
  Ferroelectr. {\bf 25}, 399 (1980).
  
\bibitem{Ris71} T. Riste, E. Samuelsen, K. Otnes, and J. Feder, Solid
  State Commun. {\bf 9}, 1455 (1971).
  
\bibitem{Sha72} S. Shapiro, J. Axe, G. Shirane, and T. Riste, Phys.
  Rev. B {\bf 6}, 4332 (1972).
  
\bibitem{Wag80a} D. Wagner {\it et~al.}, Ferroelectrics {\bf {26}},
  725 (1980).
  
\bibitem{Wag80} D. Wagner {\it et~al.}, Z. Phys. B {\bf {37}}, 317
  (1980).
  
\bibitem{Schw91} F. Schwabl and U. T\"auber, Phys. Rev. B {\bf 43},
  11112 (1991).
  
\bibitem{And86a} S. Andrews, J.Phys.C {\bf 19}, 3721 (1986).
  
\bibitem{Rya86} T. Ryan, R. Nelmes, R. Cowley, and A. Gibaud, Phys.
  Rev. Lett. {\bf 56}, 2704 (1986).
  
\bibitem{Thu93} T. Thurston {\it et~al.}, Phys. Rev. Lett. {\bf 70},
  3151 (1993).
  
\bibitem{Cow96a} R. Cowley, Phys. Scr. T {\bf 66}, 24 (1996).
  
\bibitem{Neu95b} H.-B. Neumann, U. R{\"u}tt, and J. Schneider, Phys.
  Rev. B {\bf 52}, 3981 (1995).
  
\bibitem{Rue97a} U. R\"utt {\it et~al.}, Europhys. Lett. {\bf 39}, 395
  (1997).
  
\bibitem{Wan98a} S. Wang, Y. Zhu, and S. Shapiro, Phys. Rev. Lett.
  {\bf {80}}, 2370 (1998).
  
\bibitem{Hue00b} H. H{\"u}nnefeld {\it et~al.}, Phys. Rev. B (2000),
  submitted.
  
\bibitem{Neu94a} H.-B. Neumann {\it et~al.}, J. Appl. Cryst. {\bf 27},
  1030 (1994).
  
\bibitem{Rue95d} U. R\"utt, H.-B. Neumann, H. Poulsen, and J.
  Schneider, J. Appl. Cryst. {\bf 28}, 729 (1995).
    
  \bibitem{Edw75} S. Edwards and P. Anderson, J. Phys. F {\bf 5}, 965
    (1975).
    
  \bibitem{Mez87} M. M{\'e}zard, G. Parisi, and M. Virasoro, {\em
      ``{Spin Glass Theory and Beyond}''} (World Scientific,
    Singapore, 1987).
    
  \bibitem{Dot95a} V. Dotsenko, A. Harris, D. Sherrington, and R.
    Stinchcombe, J. Phys. A: Math.  Gen. {\bf 28}, 3093 (1995).
  
  \bibitem{Nar99} R. Narayanan, T. Vojta, D. Belitz, and T.
    Kirkpatrick, Phys. Rev B {\bf 60}, 10150 (1999).
    
  \bibitem{Mar99} E. Marinari {\it et~al.}, {J. Stat. Phys.} {\bf 98},
    973 (2000).
    
  \bibitem{Bou98} R. Bouchard {\it et~al.}, J. Synchrotron Rad. {\bf
      5}, 90 (1998).
    
  \bibitem{Hir95a} K. Hirota {\it et~al.}, Phys. Rev. B. {\bf 52},
    13195 (1995).
    
  \bibitem{Sche76} H. Scheel, J. Bednorz, and P. Dill, Ferroelctrics
    {\bf 13}, 507 (1976).
    
  \bibitem{Shi69} G. Shirane and Y. Yamada, Phys. Rev. {\bf 177}, 858
    (1969).
    
  \bibitem{Dar76} C. Darlington and O. D.A., J. Phys. C {\bf 9}, 3561
    (1976).
    
  \bibitem{Den95a} I. Denk, W. M\"unch, and J. Maier, J. Am. Ceram.
    Soc.  {\bf 78}, 3265 (1995).
  
  \bibitem{Fle96} J. Fleig, F. Noll, and J. Maier, Ber. Bunsenges.
    Phys.  Chem. {\bf 100}, 607 (1996).

  \bibitem{Rod99} S. Rodewald, J. Fleig, and J. Maier, J. Europ. 
    Ceram. Soc. {\bf 19}, 797 (1999).

  \bibitem{Mue71} K. M\"uller and W. Berlinger, Phys. Rev. Lett. {\bf
      26}, 13 (1971).
    
  \bibitem{Has78} J. Hastings, S. Shapiro, and B. Frazer, Phys. Rev.
    Lett. {\bf 40}, 237 (1978).
  
  \bibitem{Bae78} D. B\"auerle and W. Rehwald, Sol.State Com. {\bf
      27}, 1343 (1978).
    
  \bibitem{LeG80} J. LeGuillou and J. Zinn-Justin, Phys. Rev. B {\bf
      21}, 3976 (1980).
    
  \bibitem{Fis64} M. Fisher, J. Math. Phys. {\bf {5}}, 944 (1964).
    
  \bibitem{Sch78} H. Schmidt and F. Schwabl, Z. Phys. B {\bf 30}, 197
    (1978).
    
  \bibitem{Sat85} M. Sato {\it et~al.}, Phase Trans. {\bf 5}, 207
    (1985).
    
  \bibitem{Lev59} A. Levanyuk, Sov. Phys. JEPT {\bf 36}, 571 (1959).
    
  \bibitem{Gin60} V. Ginzburg, Sov. Phys. Sol. State {\bf 2}, 1824
    (1960).
    
  \bibitem{Dot95b} V. Dotsenko and D. Feldman, J. Phys. A: Math. Gen.
    {\bf 28}, 5183 (1995).
  
  \bibitem{Kau00} B. Kaufmann, Ph.D. thesis, Technische
    Universit{\"a}t M{\"u}nchen, 2000.
    
  \bibitem{Kor96} A. Korzhenevskii, K. Herrmanns, and W. Schirmacher,
    Phys. Rev. B. {\bf 53}, 14834 (1996).
    
  \bibitem{Kau99} B. Kaufmann, F. Schwabl, and U. T\"auber, Phys. Rev.
    B {\bf {59}}, 11226 (1999).
    
  \bibitem{Tae92} U. T\"auber and F. Schwabl, Phys. Rev. B {\bf 46},
    3337 (1992).

\end{thebibliography}

\newpage
\begin{table}[!ht]
    \begin{tabular}[ht]{l|l|l|l}
Sample \#&Growth technique&sample preparation&sample colour\\
\hline
\hline
I&\bf{Floatzone grown}&-&Brownish, transp.\\
\hline
II&\bf{Flux grown}&etched (89\%\ H$_3$PO$_4$, 1h, 160$^\circ $C)&Brownish, transp.\\
\hline
&\bf{Verneuil grown}&&\\
III&Oxidized&(48h, O$_2$-atm., 1bar, 1000$^\circ $C)&Rose, transp.\\
IV&As grown &-&transparent\\
V&Reduced&(5h, H$_2$-atm., 1bar, 1250$^\circ $C)&Black
    \end{tabular}
    \caption {Specifications of the samples investigated. Specimens grown by three different techniques (bold type) were available. \label{tab:prep}}
\end{table}
\begin{table}
    \begin{tabular}[h]{l|c|c|c|c|c|}
&\tc&n [cm$^{-3}$]&d [\AA]&$\frac{1}{d}$ [10$^{-3}$\AA$^{-1}$]&$\tau_d$\\
\hline
\hline
{\bf Floatzone grown}&98.8(1)&6.1(2)$\times$10$^{18}$&55(1)& 18.2(2)&$\sim$ 0.115\\
\hline
{\bf Flux grown}&102.6(2)&2.8(2)$\times$10$^{18}$&71(2)&14.2(3)&$\sim$ 0.035\\
\hline
{\bf Verneuil grown}&&&&&\\
Oxidized&105.7(1)& 7.4(2)$\times$10$^{16}$&238(2)& 4.2(1) & $\sim$ 0.014\\
As grown&105.8(1)& 7.6(2)$\times$10$^{16}$&236(2)& 4.2(1) & $\sim$ 0.025\\
Reduced &101.0(1)& 1.7(1)$\times$10$^{19}$& 39(1)&25.7(5) & $\sim$ 0.20
    \end{tabular}
 \caption {Critical temperatures \tc{} and defect concentrations $n$ of the different samples. $d=n^{-1/3}$ is the mean distance of defects, $\tau_d$ is taken from the measurements of the inverse correlation length ($\kappa(\tau_d)=d^{-1}$).  \label{tab:samples}}
\end{table}
\newpage
\newpage
\newpage
\newpage
\newpage
\newpage
%
%
%
\end{document}